# Electrokinetic Power Harvesting from Wet Textile

**Sankha Shuvra Das, Vinay Manaswi Pedireddi, Aditya Bandopadhyay, Partha Saha and Suman Chakraborty***

Department of Mechanical Engineering, Indian Institute of Technology Kharagpur, Kharagpur 721302, INDIA.

*email: suman@mech.iitkgp.ernet.in

**Developing low-weight, frugal and sustainable power sources for resource-limited settings appears to be a challenging proposition for the advancement of next-generation sensing devices and beyond. Here, we report the use of centimeter-sized simple wet fabric pieces for electrical power generation, by deploying the interplay of a spontaneously induced ionic motion across fabric channels due to capillary action and simultaneous water evaporation by drawing thermal energy from the ambient. Unlike other reported devices with similar functionalities, our arrangement does not necessitate any input mechanical energy or complex topographical structures to be embedded in the substrate. A single device is capable of generating a sustainable open circuit potential up to ~700 mV. This suffices establishing an inherent capability of functionalizing self-power electronic devices, and also to be potentially harnessed for enhanced power generation with feasible up-scaling.**

## Introduction

In the milieu of global warming and energy crisis, frugal yet sustainable and renewable energy resources are critical to the advancement of human civilization[1–4]. Generating sustainable and stable electricity from mundane and natural sources without necessitating mechanized inputs and sophisticated setups triggers wishful desires and compelling challenges, simultaneously[5–9]. Natural drying of wet clothes is common in daily lives, which is a ubiquitous process ideally suited for harvesting thermal energy from the incipient ambient[10–12]. During this drying process, one may potentially harness the motion of salt-water solution within the fabric matrices towards creating preferential charge segregation[13], generating electrical power[14–18].

By introducing fabrication of simply-designed channels on textile pieces for achieving guided transportation of saline water and drawing analogies of the later with evaporative transport across the parts of a living plant, here we demonstrate an extremely frugal, simple and viable approach of harvesting electrical power using cellulose-based fiber cloth. In sharp contrast to other reported devices deploying specially-structured and delicately-fabricated substrates towards achieving this feat[1,8,19–26], we deploy a regular cellulose-based wearable textile as a medium for ionic motion though the interlace fibrous network by capillary action, inducing an electric potential in the process. The device design inherently exploits a large transpiration surface for achieving a sustainable salt-ion-flux migration, through natural evaporation effect. Our approach essentially deploys the surface energy, an intrinsic property of the material, to convert into electricity, in this manner. Notably, upon solar irradiance, the device can effectively generate a significantly higher potential than that in the absence of sun-light.

A single prototype fabric channel (FC) harvester (with stem width of 4 cm, stem length of 3 cm and leaf area of 7 cm by 7 cm) shows the ability to generate a maximum open circuit potential of ~700 mV. Featuring these functionalities, the device can be exploited to electrically functionalize paper- or fabric-based rapid kits for on-spot diagnostics of diseases in resource-limited settings.

# Experimental details

## Chemicals

Sodium chloride, NaCl (>99% purity) was purchased from Merk Life Science Pvt. Ltd. Fabric cloth (white color) of ~95% cotton was purchased from local garment shop. Copper sheet of ~300 µm thickness (>98%), plastic straw was purchased from local market. All the electrolyte solutions are prepared by dissolving the particular analyte at various concentrations in DI water (Millipore India, 18.2MΩ-cm).

## Device description and experimentation

FCs are prepared from a piece of commercial wearable fiber cloth, where those are cut to the required channel design using scissors. FC has three segments such as root, stem and leaf (Fig. 1a) which is physically analogous to the transport mechanism in a leaf (Fig. 1b and Supplementary Fig. 1). Field Emission Scanning Electron Micros copy (FESEM) analysis shows the stitching pattern of the fibers with crisscross orientation having periodically alternating micropores, with fiber interspacing of ~5-6 µm (Fig. 2b). The electrodes made of commercial grade copper sheet (>98% purity), with dimensions ~2 cm × 1 cm and ~300 µm thickness, are embedded at either sides of the stem section (ensuring a proper ohmic contact between FC surface and electrode). In order to obtain a rising capillary motion, the channel is inserted vertically into an electrolyte (aqueous NaCl solution) filled beaker covering the root area of FC, and leaving the leaf area exposed to the atmosphere (Fig. 1b). The bigger leaf area promotes the evaporation of water while maintaining a continuous negative pressure gradient across the channel length. To limit the evaporative loss, the entire stem section is inserted into a plastic straw (diameter ~6 mm) with length equal to that of stem length. Prior to the experimentation, FC is cleaned with DI water followed by drying.

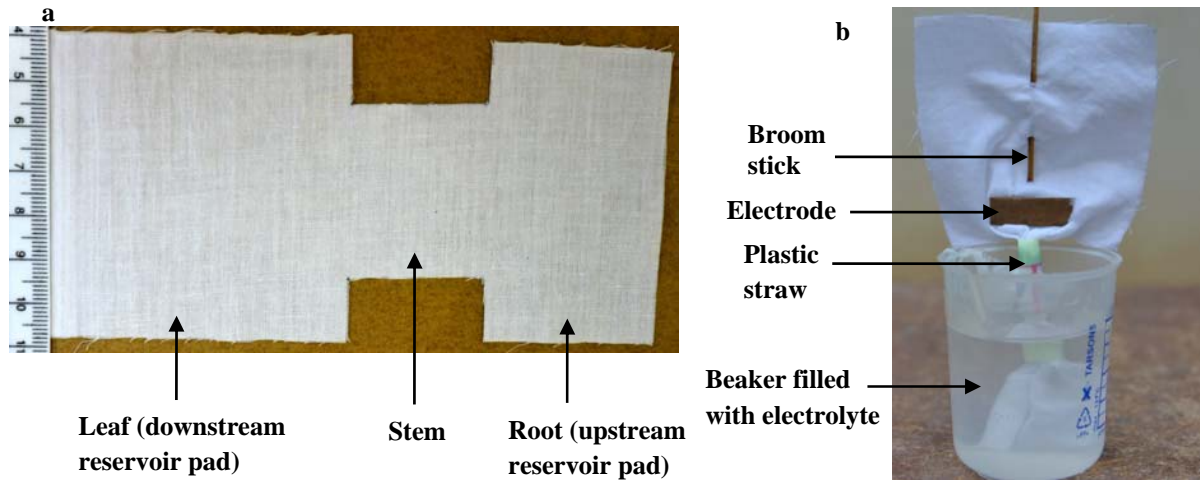

**Figure 1| Design of a fabric-based channel (FC). a,** Photograph of a FC prepared from a commercial grade fiber cloth. **b,** Complete FC based electrokinetic power generator; demonstrating its various segments.

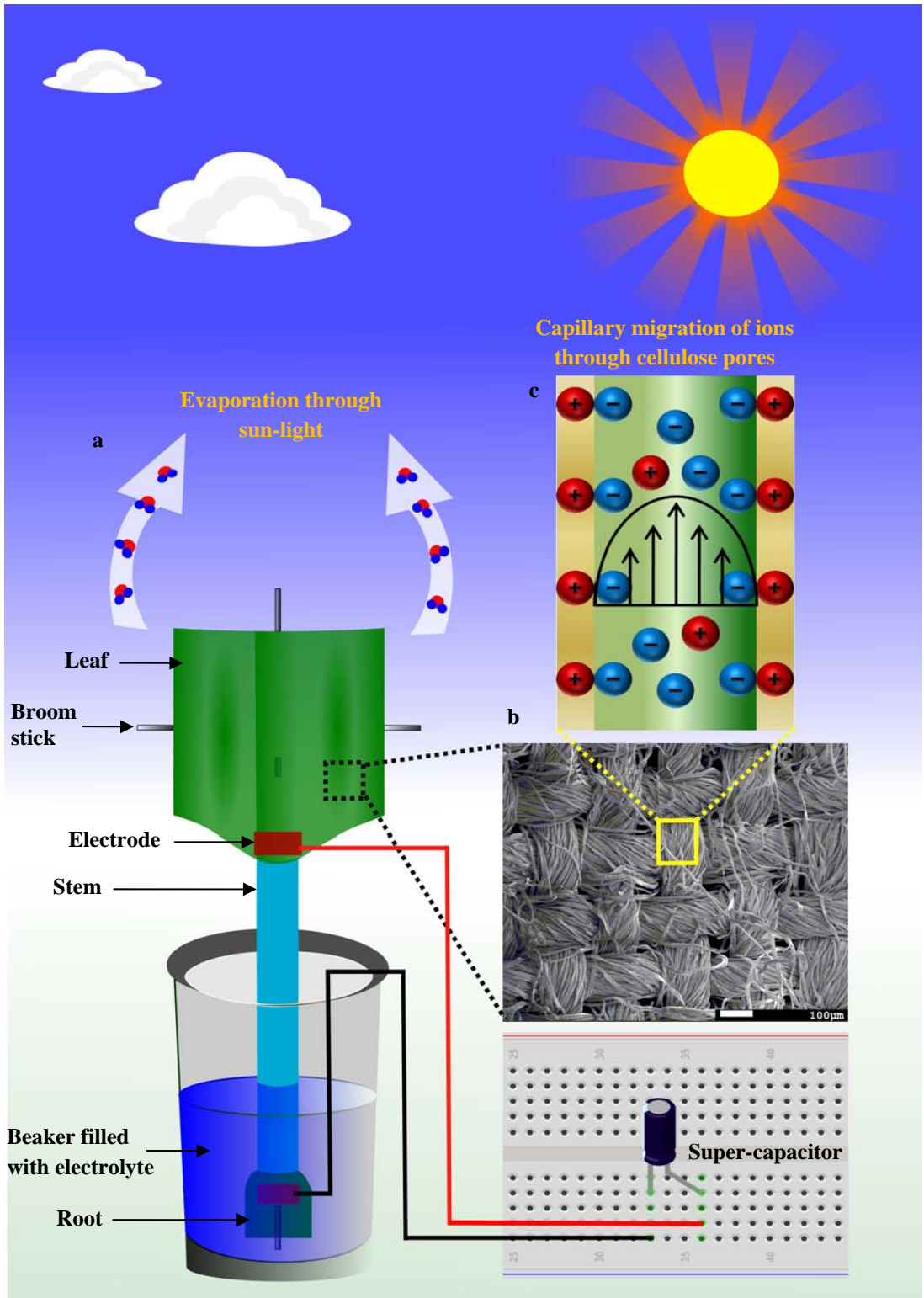

**Figure 2| Capillarity-coupled-evaporation driven FC based electrokinetic power generator. a,** Schematic representation of the experimental setup for measuring induced potential. **b,** FESEM image of FC showing the stitching pattern and crisscross orientation of fibers. **c,** Capillary transport of ions through individual pore, showing specific ion distribution inside the pore.

The schematic representation of the complete experimental setup used for capillarity-coupled-evaporation induced streaming potential measurement is illustrated in Fig. 2. In order to measure the induced electric potential and the corresponding short-circuit current, nanovoltmeter (Keithley 2182A) and picoammeter (Keithley 4185) are connected in series and parallel to the copper electrodes, respectively. The high-end of the nanovoltmeter probe is connected to the downstream electrode, whereas the low-end of the nanovoltmeter probe is connected to the upstream electrode, and we term it as straight polarity. Before the measurements, the channels are primed with the electrolyte solution for almost 1 hour. The entire set of experiments are performed under the sun-light of hot-summer days (from March to June, 2018 at Kharagpur, India; latitude 22.3460° N, longitude 87.2320° E), where the temperature, T and relative humidity, RH fluctuate between ~32-40 °C and ~60-85%. Hence, to maintain uniform experimental conditions such as T ~35 °C and RH ~60%, the experiments are further conducted inside a controlled temperature and humidity test chamber (Labard Instrument Pvt. Ltd.). All the tests are performed thrice to check the repeatability of the measurements. The following are the FC dimensions; root area ($A_R$) ~7×3 cm$^2$, stem width ($W_S$) ~4 cm, stem length ($L_s$) ~3 cm and leaf area ($A_L$) ~7×7 cm$^2$ (optimized FC dimensions; and discussed in the subsequent section), followed during the entire course of experimentation.

**Characterization**

The surface morphology of fabric cloth, structures of the fiber, pore size and distributions are characterized by Field Emission Scanning Electron Microscope, FESEM (JEOL, Model: JSM-7800 F). The electrical impedance of FC in the presence of 1 mM NaCl solution is measured using Potentiostat (GAMRY, Reference 600). Zeta-potential is measured using Electrokinetic analyzer for solid surface (Anton Paar GmbH, Model: SurPASS 3).

## Results and discussion

**Optimization of FC dimensions**

As the liquid imbibes through the porous and tortuous network of the cellulose fibers, an electrical potential is induced between two copper electrodes which gradually rises, and thereafter reaches a maximum value followed by a steady state upon absolute saturation of leaf area, $A_L$. The ions (such as Na$^+$, Cl$^-$, H$^+$ and OH$^-$) in the electrolyte solution (1 mM NaCl) creates an interfacial charge layer (also known as electrical double layer; EDL) upon interaction with negatively charged free surface groups (such as –COOH, –OH) of cellulose fabrics[25,27]. The counter-ions (Na$^+$, H$^+$) thus preferentially migrate uphill through the mobile layer of the EDL under a capillary driven pressure gradient, which eventually creates a charge polarization, leading to induced streaming potential. Initial fluctuations in the readings can be attributed from the fact of rapid rise of the capillary front which eventually equilibrates with the pore saturation.

We have performed sets of comprehensive experiments in order to understand the effect of FC dimensions (such as $L_S$, $A_L$ and $W_S$) on the induced streaming potential. Initially, the stem length (or the capillary length), $L_S$ is varied, keeping $W_S$ ~3 cm and $A_L$ ~5×5 cm$^2$ constant. The individual saturation voltage (or steady-state voltage), $V_{sat}$ and short-circuit current, $I_{sc}$ is recorded in three different identical test-sets, TS, which ensures the repeatability of the measurement. The measurements are performed in straight polarity mode. It is observed that as the capillary length increases, the corresponding capillary rising time also increases with the consequence of reduced $V_{sat}$ (Fig. 3a) and $I_{sc}$ (Supplementary Fig. 2b). The recorded saturation time for different $L_S$ such as 3 cm, 5 cm, 7 cm, 9 cm, and 11 cm are approximately 10-12 min, 18-20 min, 30-35 min, 45-50 min and 60-65 min, respectively, which further changes with the number of runs. The channel resistance, which is a function of $L_s$ (Supplementary Fig. 3; where Nyquist plot shows the variation of channel impedance with respect to stem length), increases with the channel length. This further impedes the transport of ions, and reduces $V_{sat}$ and the corresponding $I_{sc}$. Thus, to estimate the induced output power against different $L_s$, a commercial capacitor of ~0.1 mF is connected to the circuit. The output power is calculated by following the relation, $P_{output} = 0.5CV_c^2/t_c$; where $C$ is the capacitance value, $V_c$ is the capacitor charging voltage and $t_c$ is the capacitor charging time. We observed that $P_{output}$ for all these three sets decreases with increase in $L_s$ (Supplementary Fig. 2a). Thus, a lower FC length ($L_s$ ~3 cm) delivers a stable open-circuit potential of ~350 mV with $I_{sc}$ of ~1.75 µA and $P_{output}$ of ~110 nW.

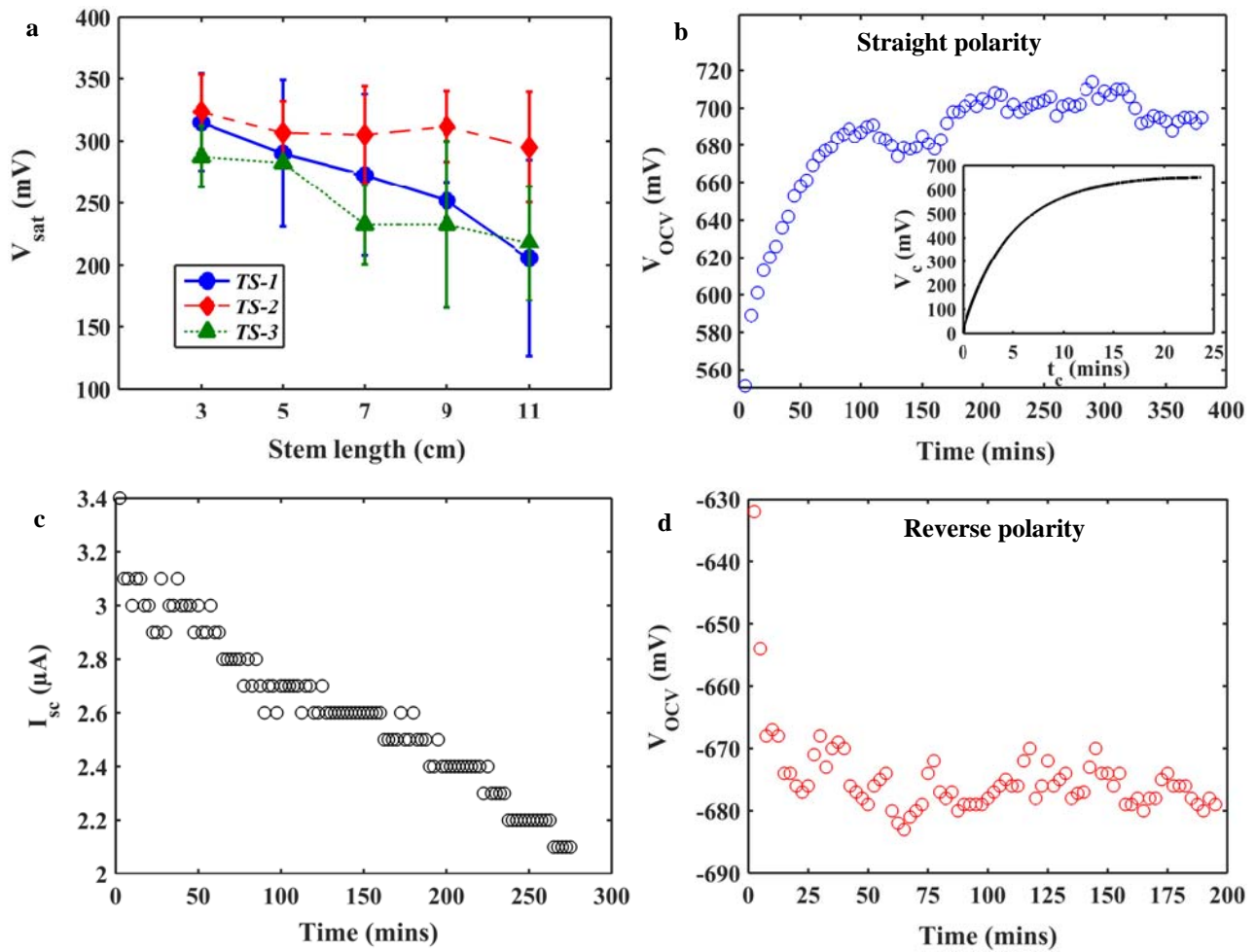

**Figure 3| Optimization of FC dimension. a,** Saturation voltage with respect to stem length of FC. **b**, Temporal evolution of induced electric potential for OFC connected in straight polarity, inset shows time required for charging a capacitor of ~0.1 mF. **c**, Temporal evolution of short-circuit current for OFC. **d**, Temporal evolution of induced potential for OFC connected in reverse polarity investigated under environmental conditions such as T ~32-35 °C and RH ~65-75%. All the measurements are performed in presence of 1 mM NaCl solution.

To further explore the effect of leaf area, $A_L$ and stem width, $W_S$ on induced potential, $P_{output}$ as well as $I_{sc}$ is measured against different $A_L$ (Supplementary Fig. 2c and 2d) and $W_S$ (Supplementary Fig. 2e and 2f) respectively. The bigger leaf area corresponds to a larger transpiration surface which essentially promotes the evaporation rate of water, resulting in a faster ion migration, and thereby induces a higher output power. The short-circuit current, $I_{sc}$ increases initially with $A_L$. However, it decreases after a certain critical value (such as ~5×5 cm$^2$), which can be attributed to increase in channel impedance only. On the other hand, the parameter $W_S$, an intrinsic function of the fiber density, effectively can stimulate a higher $P_{output}$ upto certain critical value (such as $W_S$ ~4 cm). The increase in $W_S$ reduces the channel hydraulic resistance which eventually increases $I_{sc}$. However, further rise in $W_S$, may amplify the channel impedance that leads to reduction in $P_{output}$. Based on the overall parametric study, an optimized FC (OFC) dimensions that deliver maximum $P_{output}$ become: $L_s$ ~3 cm, $A_L$ ~7×7 cm$^2$ and $W_S$ ~4 cm. The OFC is able to reliably generate a stable $V_{OCV}$ of ~580-700 mV (Fig. 3b; inset shows the charging characteristics of a capacitor) with $I_{sc}$ of ~2.1-3.4 µA (Fig. 3c) and $P_{output}$ of ~153.5 nW (see inset of Fig. 3b), maintained for ~8 hours of experiment under the sunlight, with environmental conditions such as T ~32-35 °C and RH ~65-75%. In order to further confirm the effect of EDL on induced streaming potential, we change the nanovoltmeter probes in reverse connection mode, where $V_{OCV}$ changes its polarity and thereby confirms the effect of negative surface charge on the induced electric potential (Fig. 3d).

**Device performance**

In order to analyze the role of temperature and associated evaporative flux on device performance, we perform a day-night cyclic test which consists of an uninterrupted measurement for 3 days (Fig. 4a). The maximum $V_{OCV}$ is recorded as ~550-600 mV at peak-day hours (12:30 pm-2 pm), where the effect of solar-heat and water evaporation is observed to be significant. The device performance deteriorates gradually from its peak value as the day-light intensity reduces and reaches to a minimum value of ~180-200 mV during the mid-night (1 am-3 am), where both these effects are quite less. The device performance improves again in the next day as the sun starts rising and subsequently reaches to a maximum value during the peak-day hours (almost equivalent to previous day's V$_{OCV}$ value) and thus the cycle continues. It is worth noting that the overall performance of the device is almost same at the end of the cyclic test (i.e. after 72 hours). Thus, to ensure evaporation as the prime source of induced potential, we expose the leaf area to a hot-air (T ~65°C) blowing at a speed of ~2.5 m/s. It enhances the water evaporation rate from the leaf area, leading to gradual increase in voltage maximum up to ~585 mV (Fig. 4b). The induced potential drops to ~550 mV as the hot-air blower stops, and thus represents a significant correlation between the device performance and the rate of water evaporation.

On the other hand, to appreciate the performance of the device for long-term usages, and hence to determine the robustness of the device, we conduct an experiment (exposed to day-light only) with three different channels, subjected to three different conditions for more than 30 days' (Fig. 4c). The first channel is continuously exposed to electrolyte solution (abbreviated as $TS_C$) for the total 30 days' duration of experiment; the second channel is cleaned using DI water and dried after each day's measurement (abbreviated as $TS_{CD}$); and the third channel is only dried after each day's measurement under room conditions (abbreviated as $TS_D$). The readings are taken at different day-time intervals including the peak-day hours. It is observed that the device performance remains almost constant throughout the entire 30 days' span of experiment for all these three sets (Fig. 4c), with minor fluctuations in the readings, which can be ascribed from the fact of change in ambient conditions. Hence, the device can be treated as a 'greener power-plant' and can be used as a source of continuous power supply (at the expense of zero external energy consumption).

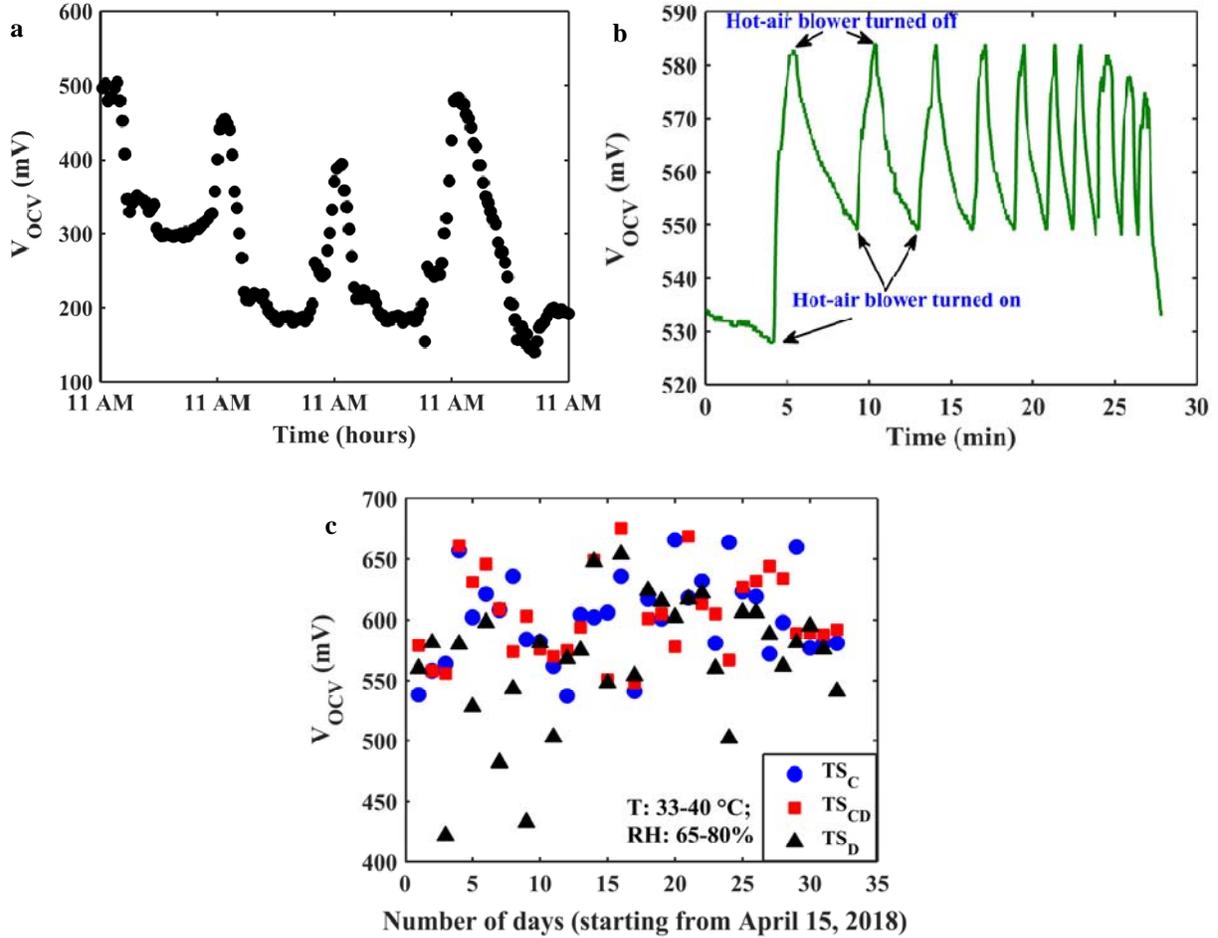

**Figure 4| Device performance in different ambient conditions. a,** Day-night cyclic test, showing the variation of induced potential of 3 days' uninterrupted measurement. **b,** Effect of induced potential on periodic application of hot-air blower. **c,** Performance of OFC subjected to three different conditions against 30 days' measurement (measurement started from April 15, 2018).

## Conclusions

To summarize, we have shown that water evaporation from a centimeter-sized piece of wet cloth containing frugally-cut fabric channels can generate electrical power. Drawing analogies with evaporative transport in living plants and harnessing the consequent preferential migration of ions towards developing an electrical potential as mediated by capillary action, we have demonstrated that ordinary cellulose-based wet textile, may be capable enough for achieving this remarkable feat. As compared to previously reported methods of energy harvesting from complex resources, the electricity generation occurs in natural ambience, directly converting the abundantly available thermal energy into electrical power. Further, in contrast to classical streaming potential generated by an applied pressure gradient or other external pumping resources, here the intrinsic surface energy of the fabric is used to drive the ionic current. Most importantly, our method paves the way of deploying regular fabric pieces as the sources of energy, with no special topographical manipulation of the cloth surface being demanded. Thus, the device does not necessitate any extensive fabrication protocol, unlike some recently reported evaporation driven energy harvesting devices. Finally, in a hot and dry environment, the natural evaporative transport gets spontaneously enhanced, so that the flow-induced electrical potential can be maximized. The device, thus, may turn out to be extremely effective in geographically warm and dry regions of the earth. Our results reveal that a single fabric channel can stably deliver a potential of ~700 mV in ambient conditions. This eventually culminates into a utilitarian paradigm of low-cost power harvesting in extreme rural settings.

# Supplementary Information

## Electrokinetic Power Harvesting from Wet Textile

**Sankha Shuvra Das, Vinay Manaswi Pedireddi, Aditya Bandopadhyay, Partha Saha and Suman Chakraborty***

Department of Mechanical Engineering, Indian Institute of Technology Kharagpur, Kharagpur 721302, INDIA.

*email: suman@mech.iitkgp.ernet.in*

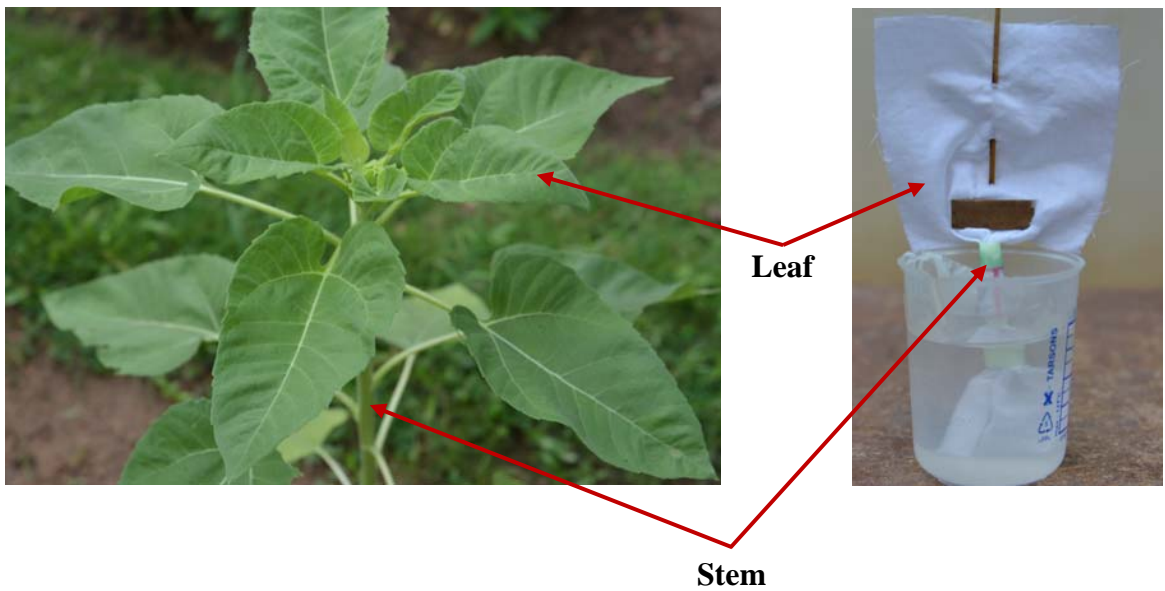

**Leaf**

**Stem**

**Supplementary Figure 1| Analogy of transport mechanism of a tree.** Photograph representing the analogy of the various segment of a tree and fabric-based channel. Bigger leaf area promotes the evaporation rate.

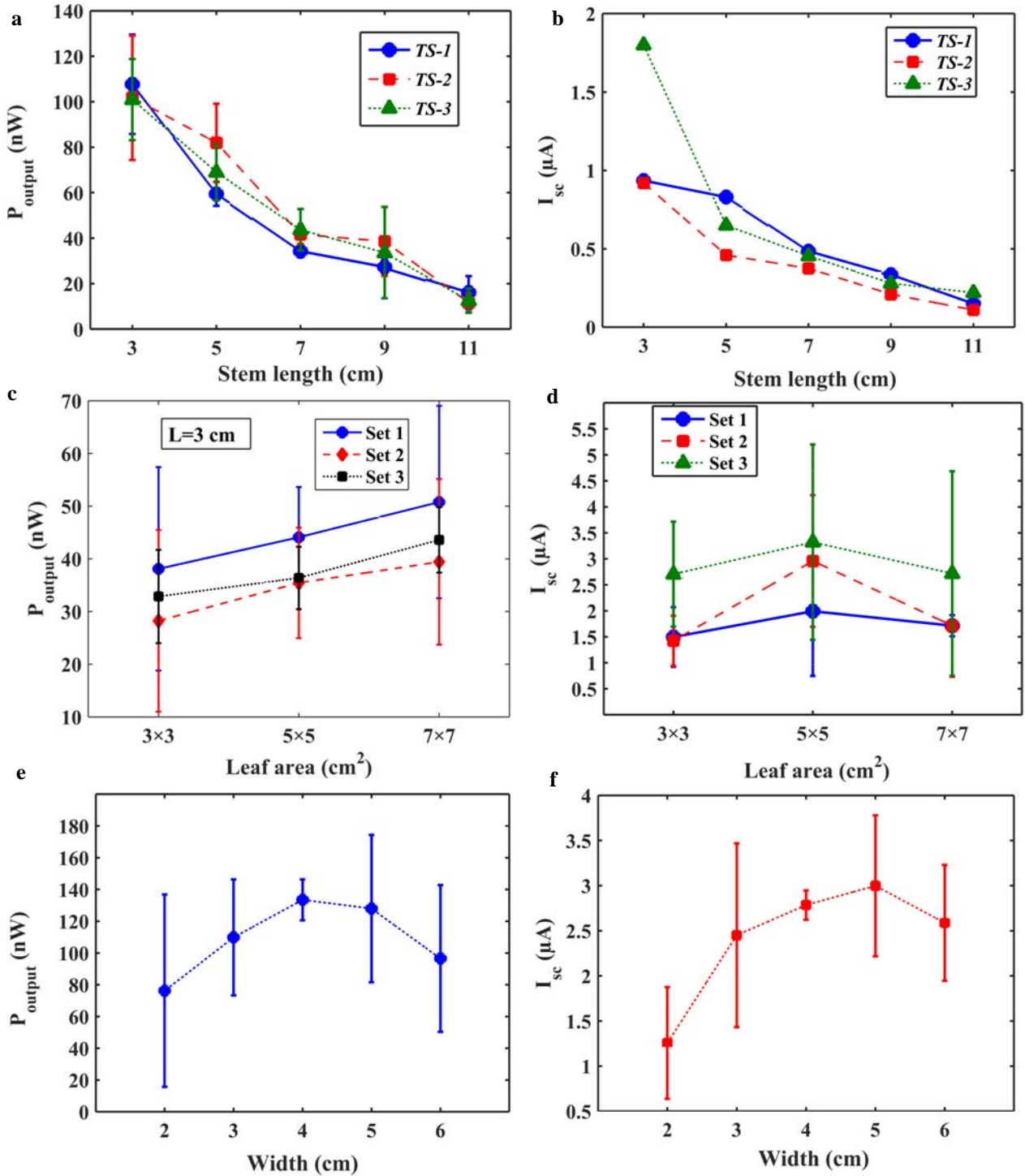

**Supplementary Figure 2| Parametric optimization of FC.** (a) Output power, and (b) Short-circuit current with respect to stem length (with constant stem width of 3 cm and leaf area of 5×5 cm) of FC. (c) Output power, and (d) Short-circuit current with respect to leaf area (for optimized stem length of 3 cm and constant stem width of 3 cm) of FC. (e) Output power, and (f) Short-circuit current with respect to stem width (for optimized stem length of 3cm and optimized leaf area of 7×7 cm) of FC.

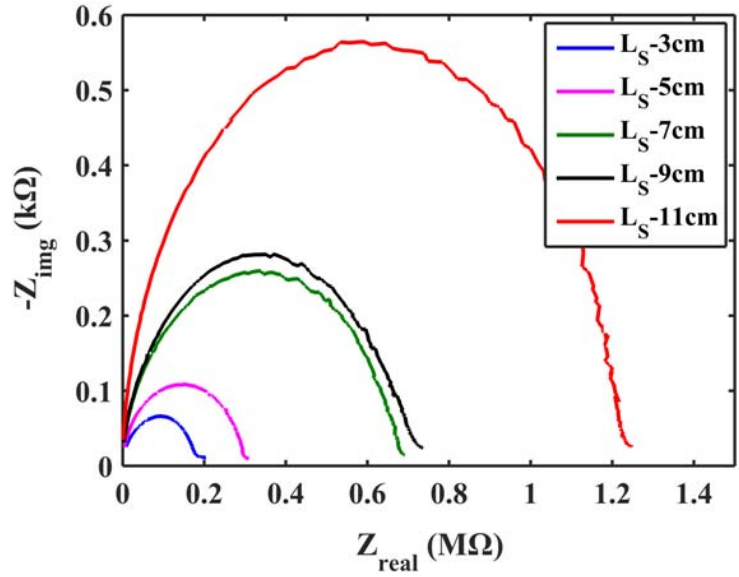

**Supplementary Figure 3| Impedance measurement of FC.** Nyquist plot represents the variation of impedance with respect to stem length of FC for 1 mM NaCl solution.

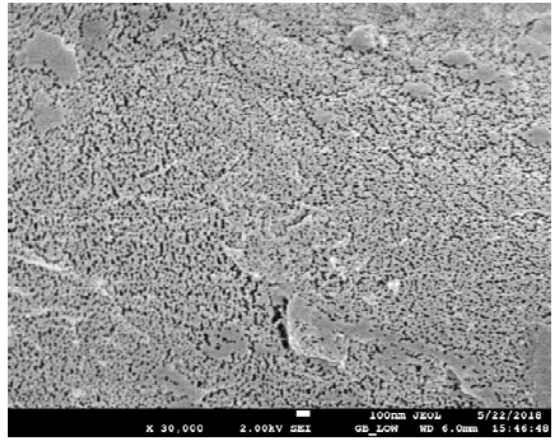

**Supplementary Figure 4| Pore size and distribution.** FESEM image (at 30,000× magnification) of single fibre thread showing the pore size and pore distribution. The Scalebar on the panel represents 100 nm.